\documentclass[%
 reprint,
 amsmath,amssymb,
 aps,
]{revtex4-2}

\RequirePackage{XCharter}
\RequirePackage[xcharter,bigdelims,vvarbb]{newtxmath}
\RequirePackage[scaled=1.1]{zlmtt}
\usepackage[upgreek, LGRgreek]{mathastext}

\usepackage{color,soul}
\usepackage{hyperref}% add hypertext capabilities
\hypersetup{
    colorlinks=true, citecolor = blue, urlcolor = blue, linkcolor = blue,
    }
\usepackage{graphicx}% Include figure files
\usepackage{dcolumn}% Align table columns on decimal point

\usepackage{float}
\usepackage{placeins}

\usepackage{parskip}
\begin{document}

\preprint{APS/123-QED}

% ---------------------------------------------------------------------------------
% ------------------------------- Title & Authors ---------------------------------
% ---------------------------------------------------------------------------------

\title{OpenDosimeter: Open Hardware Personal X-ray Dosimeter}
\author{Norah Ger\textsuperscript{1}}%
\author{Alice Ku\textsuperscript{2}}%
\author{Jasmyn Lopez\textsuperscript{2}}%
\author{N. Robert Bennett\textsuperscript{2}}%
\author{Jia Wang\textsuperscript{3}}%
\author{Grace Ateka\textsuperscript{4}}%
\author{Enoch Anyenda\textsuperscript{5}}%
\author{Matthias Rosezky\textsuperscript{6}}%
\author{Pamela Kilavi\textsuperscript{7}}%
\author{Adam S. Wang\textsuperscript{2}}%
\author{Kian Shaker\textsuperscript{2}}%
 \email{Corresponding author: kiansd@kth.se\newline Present adress: Department of Applied Physics, KTH Royal Institute of Technology, Stockholm, Sweden}

\affiliation{%
\hfill \break
 \textsuperscript{1}\mbox{Mama Lucy Kibaki Hospital, Nairobi County, Kenya} \\
 \textsuperscript{2}\mbox{Department of Radiology, Stanford University, California, US}\\
 \textsuperscript{3}\mbox{Department of Environmental Health and Safety, Stanford University, California, US} \\
  \textsuperscript{4}\mbox{Secondary Standards Dosimetry Laboratory, Kenya Bureau of Standards, Nairobi, Kenya} \\
 \textsuperscript{5}\mbox{Department of Diagnostic Imaging and Radiation Medicine, University of Nairobi, Nairobi, Kenya} \\
 \textsuperscript{6}\mbox{NuclearPhoenix, Vienna, Austria} \\
 \textsuperscript{7}\mbox{School of Computing and Engineering Sciences, Strathmore University, Nairobi, Kenya} \\
}%

\date{\today}

\begin{abstract}
We present OpenDosimeter (\href{http://www.opendosimeter.org}{www.opendosimeter.org}), an open hardware solution for real-time personal X-ray dose monitoring based on a scintillation counter. Using an X-ray sensor assembly (LYSO + SiPM) on a custom board powered by a Raspberry Pi Pico, OpenDosimeter provides real-time feedback (1 Hz), data logging (10 hours), and battery-powered operation. One of the core innovations is that we calibrate the device using $^{241}$Am found in ionization smoke detectors. Specifically, we use the $\gamma$-emissions to spectrally calibrate the dosimeter, then calculate the effective dose from X-ray exposure by compensating for the scintillator absorption efficiency and applying energy-to-dose coefficients derived from tabulated data in the ICRP 116 publication. We demonstrate that this transparent approach enables real-time dose rate readings with a linear response between 0.1--1000 µSv/h at $\pm$25\% accuracy, tested for energies up to 120 keV. The maximum dose rate readings are limited by pile-up effects when approaching count rate saturation ($\sim$77 kcps at $\sim$13 µs average pulse processing time). The total component cost for making an OpenDosimeter is <\$100, which, combined with its open design (both hardware and software), enables cost-effective local reproducibility on a global scale. Through a student workshop, we also demonstrate its effectiveness as an educational and capacity-building tool. This paper complements the open-source documentation by explaining the underlying technology, the algorithm for dose calculation, and areas for future improvement.
\end{abstract}

\maketitle

\section*{\label{sec:intro} Introduction}
\vspace{-2mm}
Monitoring occupational X-ray exposure is critical for ensuring the short- and long-term safety of radiation workers. To minimize health risks, regulations set dose limits for these workers, such as a yearly effective dose below 20 mSv (averaged over five consecutive years) and not exceeding 50 mSv in any single year \cite{IAEA2018}. However, while there is limited data on the global availability, many workers in low-resource settings have little or no access to personal dosimeters. Traditional dosimeters can be prohibitively expensive and require complicated logistics around calibration, readout, and reporting. All of this leads to radiation workers having restricted insight into their own exposure levels.

Personal dosimeters are categorized as either active or passive devices. On the passive side, dosimeters based on thermoluminescence (TLDs) or optically stimulated luminescence (OSL) allow for cost-effective scalability and are often offered as a subscription service by an external party. However, individuals wearing these dosimeters receive infrequent feedback on their radiation exposure, typically only monthly or quarterly. In addition, the logistics of collecting dosimeters and sending them to remote facilities for reading is challenging. Moreover, the lack of immediate feedback hinders the individual in maintaining effective radiation safety practices.

On the active side, there are real-time electronic personal dosimeters. An example is the RaySafe i3 marketed towards interventional radiology procedures where dose exposure to the operators can be high, which makes direct feedback critical. Unfortunately, these active devices are often prohibitively expensive (>\$1000) for scaling to larger groups of radiation workers.

Numerous open hardware projects for radiation detection are publicly available, including those using Geiger--Müller tubes (e.g., RadiationD-v1.1 \cite{CAJOE2023}, uRADMonitor \cite{uRADMonitor}) and solid-state sensors (e.g., OpenGeiger \cite{OpenGeiger2023}, Open Gamma Detector \cite{OpenGammaDetector}, LABDOS01 \cite{LABDOS01}). However, to our knowledge, no open hardware project has yet addressed the critical challenge of accurately calculating the effective dose (in Sieverts) from external X-ray exposure.

We present OpenDosimeter (\href{http://www.opendosimeter.org}{www.opendosimeter.org}), an open hardware solution for low-cost and real-time personal radiation monitoring. This manuscript complements the open-source documentation (see our \href{https://github.com/OpenDosimeter/OpenDosimeter}{GitHub repository}) by explaining the underlying technology, the algorithms for spectral calibration and dose calculation, as well as benchmarking against a commercial active dosimeter (RaySafe i3). Lastly, we demonstrate its potential for capacity-building through a student workshop.

\begin{figure*}[ht!]
    \centering
    \includegraphics[width=\textwidth]{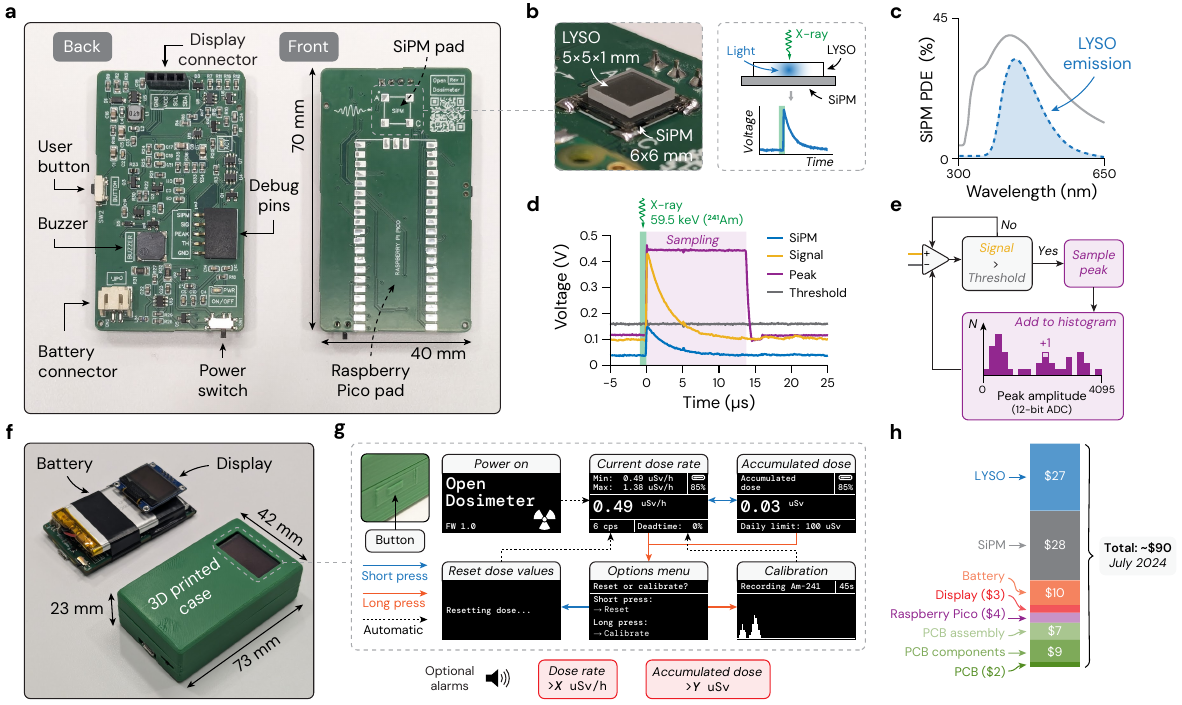}
    \caption{\textbf{Device overview.} \textbf{a}, OpenDosimeter custom board as assembled and delivered by a PCB manufacturer. \textbf{b}, X-ray sensor: LYSO crystal mounted on a SiPM. \textbf{c}, SiPM photon detection efficiency overlapping with the LYSO emission spectrum (arbitrary y-axis). \textbf{d}, Signal processing: raw SiPM pulses (blue), amplified signal (yellow), and peak signal amplitude (purple) sampled by the Raspberry Pi Pico (shaded purple); $\sim$13 $\upmu$s typical sampling time. \textbf{e}, Histogram generation from triggered sampling of peak amplitudes whenever the signal exceeds the threshold. \textbf{f}, Fully assembled OpenDosimeter, without (left) and with (right) the case. \textbf{g}, User interface: single-button navigation with short/long presses. \textbf{h}, Component cost breakdown (as of July 2024).}
    \label{fig:1}
\end{figure*}

\section*{Results}
\vspace{-2mm}
\begin{figure*}[ht!]
    \centering
    \includegraphics[width=\textwidth]{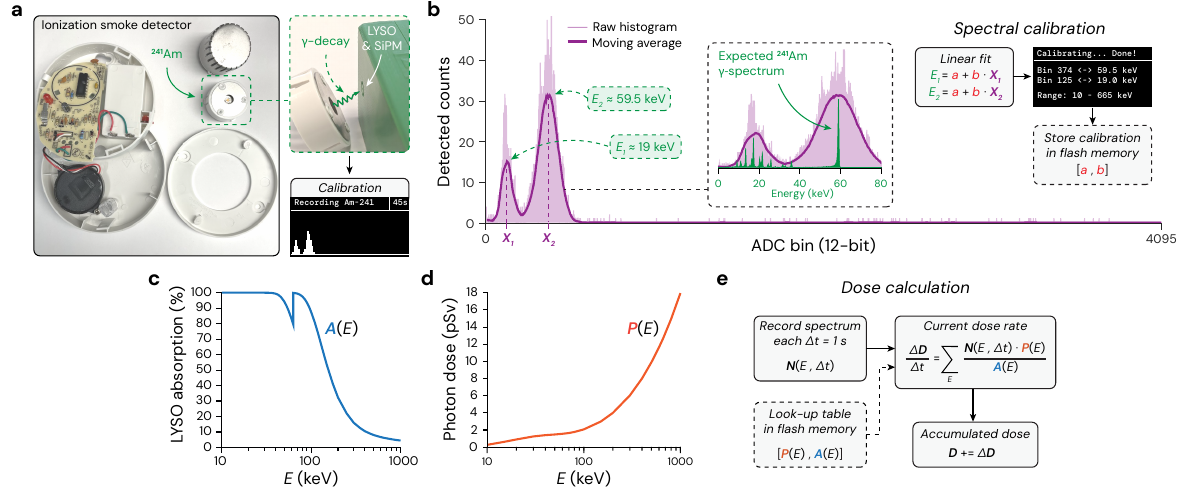}
    \caption{\textbf{Calibration and dose calculation.} \textbf{a}, Disassembled ionization smoke detector with $^{241}$Am used for spectral calibration \textbf{b}, Raw calibration histogram (light purple; shaded) and moving average (dark purple; solid) showing two peaks. Inset: zoom on the two peaks with the underlying $^{241}$Am $\gamma$-spectrum overlaid (measured with a CdTe spectrometer). The spectral calibration performs a linear fit of the two peaks with the  expected $^{241}$Am emission energies. \textbf{c}, LYSO (1 mm thick) spectral X-ray absorption efficiency. \textbf{d}, Per-photon-dose spectral coefficients, derived from ICRP 116 (Table A.1; AP direction) and scaled with the LYSO crystal area (25 mm$^2$). \textbf{e}, Algorithm for calculating the dose rate and accumulated dose.}
    \label{fig:2}
\end{figure*}

\subsection*{Device overview}
\vspace{-2mm}
The core of OpenDosimeter is the custom-designed printed circuit board (PCB, Fig.~\ref{fig:1}a). The board can be ordered, assembled, and delivered from any PCB manufacturer (see the \href{https://github.com/OpenDosimeter/OpenDosimeter}{GitHub repository}), ready for integration with the external components. The back side has analog circuitry for signal processing, connectors for peripherals (display, battery), a buzzer for sound, a button for user interaction, a power switch, and a set of breakout pins to access the analog signals for debugging. The front side has soldering pads for a Raspberry Pi Pico (the microcontroller powering the device) and for a silicon photomultiplier (SiPM). For X-ray detection, we mount a LYSO crystal on the SiPM using an optical couplant (Fig.~\ref{fig:1}b). X-ray photons absorbed by the LYSO crystal generate visible light centered in the blue region, matching the photon detection efficiency (PDE) of the SiPM (Fig.~\ref{fig:1}c). We note that the LYSO + SiPM combination is often used for $\gamma$-detection (e.g., in positron emission tomography \cite{gonzalez2016pet,melcher2000scintillation}) because of this spectral match, with care taken to subtract the weak background radioactivity of the $^{176}$Lu in the crystal ($\gamma$-emissions at 88, 202, and 307 keV)\cite{alva2018understanding}. We chose a slightly smaller LYSO crystal area (5$\times$5 mm) compared to the SiPM (6$\times$6 mm) to ensure uniform light detection by the SiPM regardless of the absorption event location in the LYSO crystal. Since we designed OpenDosimeter mainly for occupational exposure to X-rays in radiology settings, where X-ray exposure rarely exceed energies of 140 keV, we chose a 1 mm LYSO thickness as it has sufficient absorption efficiency in this energy range (see Fig.~\ref{fig:2}c).

Once X-ray photons are absorbed in the LYSO crystal, the resulting burst of blue light is detected by the SiPM, generating current then converted to a voltage pulse. SiPMs are sensitive to temperature variations (gain: -0.8\%/$\text{°C}$ for our model), so we implemented a voltage bias temperature correction circuit to mitigate this effect (now reduced to -0.25\%/$\text{°C}$). The OpenDosimeter board includes dedicated circuitry to amplify the SiPM output, which we henceforth refer to as the "signal" (Fig.~\ref{fig:1}d). This signal is then processed through an analog peak detector circuit, which holds the amplitude of the signal peak for the Raspberry Pi Pico to sample. Sampling is triggered only when the signal exceeds a threshold level determined by a comparator (Fig.~\ref{fig:1}e). We program this threshold so that it is dynamically adapted (within the 0--500 mV range) to the signal baseline using an RC-filtered pulse-width modulation (PWM) output from the Pico. When the threshold is exceeded, the Pico samples the peak amplitude and adds it to a histogram of peak amplitudes. Since these peak amplitudes are proportional to the X-ray energies, the histogram represents the detected X-ray spectrum, albeit not yet spectrally calibrated. Once calibrated, this spectrum will serve as the basis for dose rate calculations (cf. next section).

The complete OpenDosimeter is housed in a 3D-printed case measuring 73 $\times$ 42 $\times$ 23 mm (Fig.~\ref{fig:1}f), with its compact size making it wearable on the chest or belt for monitoring dose from occupational X-ray exposure.  

The device firmware implements a simple state machine for user interaction (Fig.~\ref{fig:1}g). Upon power-on, the device enters the main operational mode. Short button presses cycle through two display states; showing either the current dose rate or the accumulated dose.  The options menu is entered through a long button press, where a short press resets the dose values, and a long press brings the device to the calibration procedure (see next section). Dose values are continuously logged at 1 Hz with values for the past 10 hours can be accessed and visualized through our \href{https://opendosimeter.org/#dataLog}{Web Interface}. The interface does not require any software installation on the user end, and can be accessed simply by connecting the OpenDosimeter over USB to any computer with an internet connection. The USB connection is also used to charge the lithium-ion polymer (LiPO) battery. A battery with 1200 mAh capacity currently enables up to 20 hours of operation per full charge (power consumption typically <70 mA @ 3.7V).

Lastly,  Fig.~\ref{fig:1}h shows a cost breakdown of the OpenDosimeter. More than half of the total component cost ($\sim$\$90 as of July 2024) is from the X-ray sensor (LYSO + SiPM). In comparison, commercial dosimeters with similar functionality (real-time, battery powered, logging capabilities) are typically >\$1000 (e.g., RaySafe i3 which we benchmark against in the characterization section).

\subsection*{Calibration and dose calculation}

\vspace{-2mm}

\begin{figure*}[ht!]
    \centering
    \includegraphics[width=\textwidth]{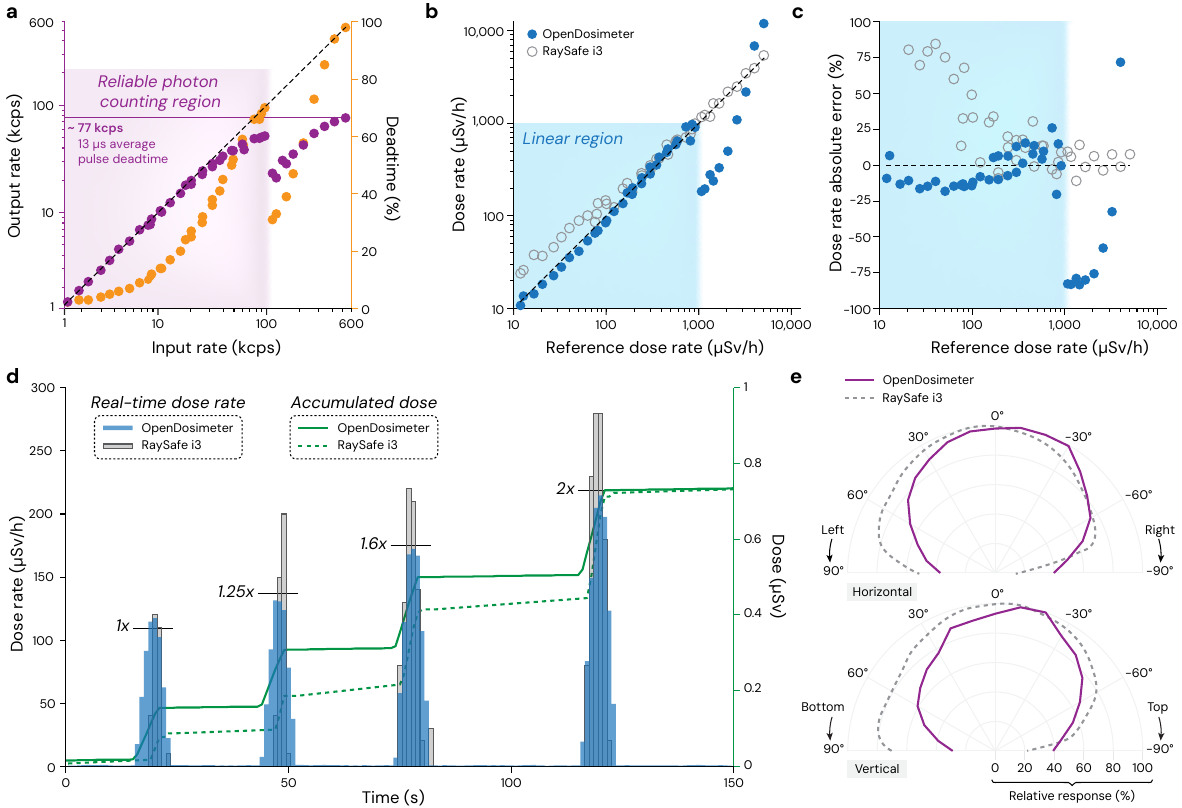}
    \caption{\textbf{Characterization and performance demonstration.} \textbf{a}, Output count rate (purple; left) and estimated deadtime (orange; right) as a function of input count rate. \textbf{b}, Output dose rate of the OpenDosimeter (blue) and the RaySafe i3 reference dosimeter (grey) plotted against the reference dose rate (RaySafe X2 survey meter). \textbf{c}, Relative errors in dose rate measurements. \textbf{d}, Demonstration of real-time dose rate performance (blue / grey) and accumulated dose (green). \textbf{e}, Relative response in the angular direction.}
    \label{fig:3}
\end{figure*}

Unlike the conventional calibration of dosimeters using the 662 keV $\gamma$-emission from $^{137}$Cs, which is typically possible only at centralized facilities, we demonstrate that the OpenDosimeter can be calibrated with any $^{241}$Am source (e.g., those found in ionization smoke detectors, cf. Fig.~\ref{fig:2}a). Specifically, we use the low-energy $\gamma$-emissions from the source to spectrally calibrate the device. Using tabulated values for the LYSO X-ray absorption efficiency and energy-to-dose conversion coefficients, we then calculate the dose rate without requiring explicit calibration against a known dose rate reference.

Figure~\ref{fig:2}b shows an example of a calibration histogram of $^{241}$Am acquired with an OpenDosimeter. The histogram recording stops when the maximum count reaches a predefined limit (e.g., 50 counts) to ensure the acquisition finishes within a reasonable time (e.g., 60 seconds). Next, we perform a moving average on the raw histogram to facilitate identification of the ADC bins corresponding to the two major peaks observed. To identify the energies of these peaks, we overlay the expected $^{241}$Am $\gamma$-emissions (Fig.~\ref{fig:2}b; inset). The peak at the higher ADC bin clearly corresponds to the 59.5 keV emission line, while the peak at the lower ADC bin likely represents the envelope of the low-energy emission lines in the range of 10-23 keV. We empirically assume that the peak of this envelope is around 19 keV, noting that an error in this assumption leads to an error in the spectral calibration and thus added uncertainty in the dose calculation (cf. next paragraph). Lastly, we assume a linear relationship between the ADC values (i.e., signal peak amplitudes) and X-ray energy, and find the corresponding linear coefficients (Fig.~\ref{fig:2}b; far right). This calibration procedure also defines the spectral range for which the device can reliably attribute the energy of detected X-ray photons. From a user perspective, this entire process is performed automatically and takes approximately 1 second once the $^{241}$Am calibration histogram has been recorded, without needing any intervention or parameter tweaking. The calibration parameters $[a, b]$ are then stored in the device flash memory and thus accessible after power cycling.

Once the device has been spectrally calibrated, we use a simple and transparent algorithm for calculating the effective dose (in Sv) from the detected X-ray spectrum. First, we correct the detected spectrum for the spectral absorption efficiency of the 1-mm thick LYSO crystal (Fig.~\ref{fig:2}c). Next, we multiply the corrected spectrum with per-photon-dose spectral coefficients (Fig.~\ref{fig:2}d) derived from tabulated data in the ICRP 116 report \cite{ICRP116}. We chose the specific coefficients assuming the device will be worn on the chest (ICRP 116; Table 1; AP direction) and scale these with the LYSO detection area ($5\times5$ mm$^2$), resulting in the graph in Fig.~\ref{fig:2}d. Data from both graphs (Fig.~\ref{fig:2}c,d) are stored in the Raspberry Pi Pico flash memory and accessible as look-up tables. Lastly, we calculate the current dose-rate at each time point ($\Delta t = 1$ s, deadtime corrected) using the detected X-ray spectrum and the look-up tables, and the accumulated dose by incrementally adding the dose per time point (Fig.~\ref{fig:2}e).

\subsection*{Characterization and performance demonstration}
\vspace{-2mm}
To characterize the OpenDosimeter performance, we conducted a series of experiments and benchmarked it against a commercial active dosimeter (RaySafe i3). Experiments were performed at X-ray energies up to 120 keV using a clinical X-ray tube to simulate the use of OpenDosimeter in a clinical radiology department for personal monitoring of occupational radiation exposure.

First, we evaluated the photon count rate performance of the OpenDosimeter (Fig.~\ref{fig:3}a). The output count rate increases linearly with the input rate up to approximately 100 kcps (thousand counts per second), defining a reliable photon-counting region (shaded purple). Beyond this point, the output becomes unreliable likely due to increasing probability of pulse pile-up effects. The maximum output count rate plateaus at $\sim$77 kcps, corresponding to a deadtime fraction approaching 100\%. The latter is estimated within the OpenDosimeter by multiplying the output count rate with the average per-pulse sampling time (cf. Fig.~\ref{fig:1}d, $\sim$13 µs) which ultimately defines the maximum output count rate. Within the reliable photon counting region, the estimated deadtime percentage can be used as a correction factor for the output count rate.

The dose rate response of the OpenDosimeter is presented in Fig.~\ref{fig:3}b, where the output is compared with the RaySafe i3 commercial dosimeter and plotted against the reference dose rate (using a RaySafe X2 survey meter). The reliable photon counting region is clearly mapped to a dose range with linear response extending up to $\sim$1 mSv/h (deadtime corrected). Beyond this range, the output becomes unreliable due to the pile-up effects nearing the count rate saturation (cf. Fig.~\ref{fig:3}a). Compared to the RaySafe i3, the OpenDosimeter performs better at low dose rates, yet the RaySafe i3 has a linear response at the higher dose rates (up to 1 Sv/h according to its datasheet).

To quantify the uncertainty of the OpenDosimeter dose-rate estimates, Fig.~\ref{fig:3}c shows the relative errors corresponding to the measurements in Fig.~\ref{fig:3}b. This shows that our dose calculation algorithm, using only $^{241}$Am for spectral calibration, results in dose rates within $\pm$25\% accuracy of the reference dose rate within its reliable region, well within range of the commercial dosimeter.

Next, we demonstrate the real-time performance of the OpenDosimeter in a dynamic radiation environment compared with the commercial dosimeter (Fig.~\ref{fig:3}d). The dosimeters were simultaneously exposed to X-rays at increasing tube currents, indicated by multiplicative factors noted as horizontal lines in the graph. Each exposure lasted 5 seconds. The plot shows both the instantaneous dose rate and the accumulated dose as measured by both devices. OpenDosimeter closely tracks the commercial device in both dose rate response and accumulated dose, showcasing the ability to accurately monitor rapidly changing radiation levels. Lastly, the angular response of the OpenDosimeter is shown in Fig.~\ref{fig:3}e with the RaySafe i3 (from datasheet) overlaid for comparison.

\subsection*{Pilot workshop for capacity-building}
\vspace{-2mm}
\begin{figure*}[ht!]
    \centering
    \includegraphics[width=\textwidth]{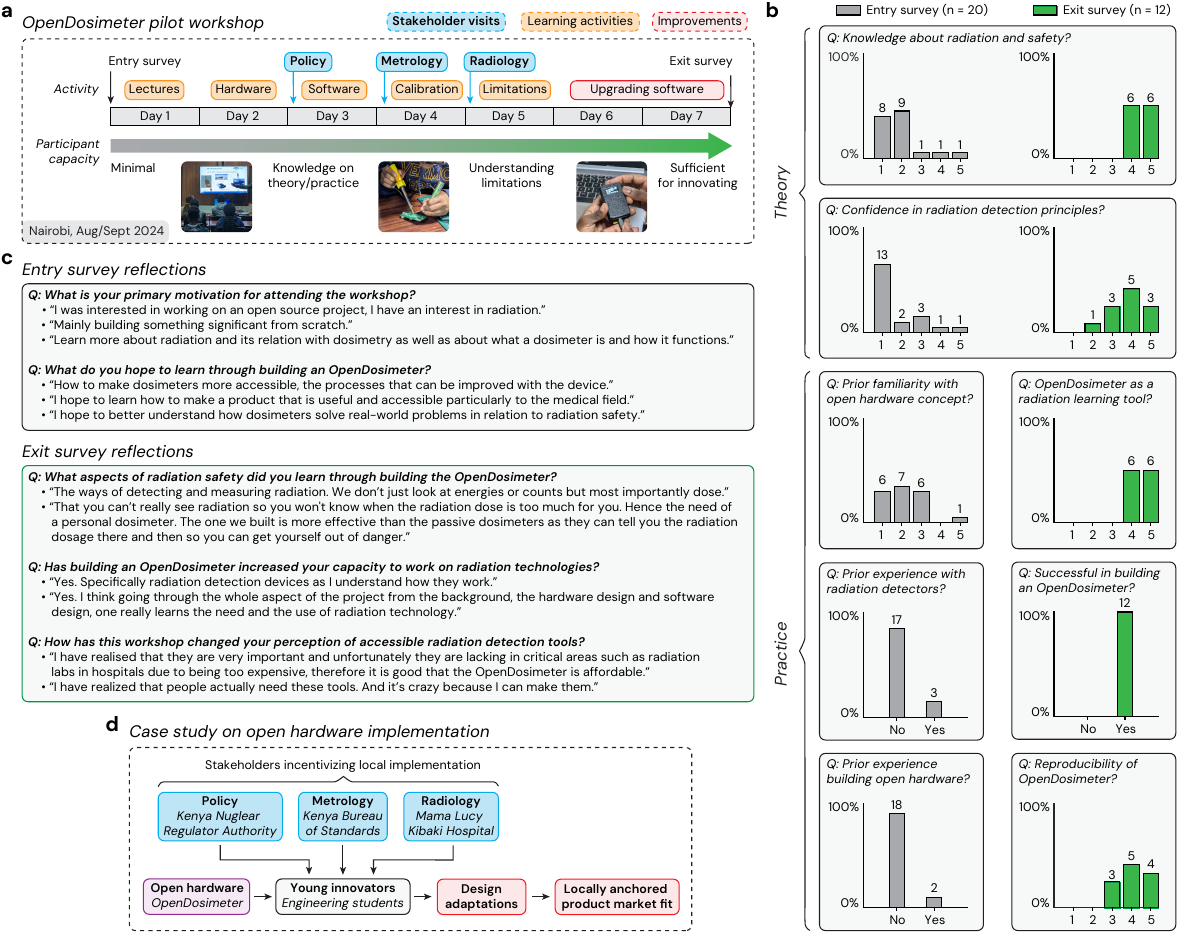}
    \caption{\textbf{OpenDosimeter as a tool for capacity-building in radiation technology.} \textbf{a}, Workshop format spanning seven days, integrating stakeholder visits, learning activities, and design improvements by the student participants. \textbf{b}, Pre- and post-workshop survey results (n=20 and n=12, respectively) showing improvements in knowledge, confidence, and practical experience, scored from 1 (low) to 5 (high). \textbf{c}, Excerpt of  reflections from entry and exit surveys highlighting participant motivations and learning outcomes (quotes lightly edited for brevity while preserving language and content). \textbf{d}, Our workshop as a case study on open hardware implementation towards locally sustainable innovation.}
    \label{fig:4}
\end{figure*}

Beyond utility as a radiation monitoring device, we evaluated OpenDosimeter as a tool for education and capacity building through a pilot workshop at Strathmore University in Nairobi, Kenya (Fig. \ref{fig:4}a). We enrolled local engineering students with minimal prior knowledge of X-ray radiation and provided them with components to build OpenDosimeters using the open-source documentation. We started the workshop with lectures on radiation safety and detection basics, then progressed to hands-on activities building dosimeters including hardware and software. We also invited local stakeholders from policy, metrology, and radiology domains to interact with the students and share their perspectives on accessible radiation monitoring.

We collected survey results at the beginning ("entry") and end ("exit") of the workshop to gain insight into the experiences of our workshop participants. The quantitative self-assessment surveys showed that by building an OpenDosimeter, our participants gained both theoretical understanding and practical capability to work with radiation technology (Fig.~\ref{fig:4}b). While most participants entered with minimal experience with both radiation detection and open hardware development, all workshop completers (n = 12, with some of the initial n = 20 participants dropping out) successfully assembled functioning OpenDosimeters. In their qualitative reflections (Fig.~\ref{fig:4}c), the students demonstrated enhanced understanding of both technical concepts and broader societal implications of accessible detection technologies for radiation safety. We found the combination of theoretical learning through lectures, hands-on experience building the OpenDosimeter hardware followed by software programming, effectively built capacity in radiation detection and safety among our participants. 

As illustrated in Fig.~\ref{fig:4}a, our participants progressed from minimal initial knowledge to achieving technical understanding sufficient for independent innovation. In the final days of the workshop, participants identified and addressed a limitation in the initial software; the need to subtract intrinsic LYSO crystal background radioactivity from the dose rate calculations. Working collaboratively, the participants implemented this improvement and we published it as a new firmware version (v1.0.1).

On a general level, our workshop can be seen as a case study demonstrating a framework for using open hardware to build local technical capacity (Fig.~\ref{fig:4}d). The engagement of local stakeholders from clinical radiology, metrology, and regulatory policy helped our student participants recognize that there exists strong incentives for local implementation. These stakeholders emphasized how locally designed and manufactured dosimeters could bridge the accessibility gap in personal radiation monitoring among radiation workers locally, showing the participants how their practical knowledge in dosimeter technology could address these needs. The alignment between open hardware and local stakeholder interests paves the way for sustainable innovation, enabling young innovators to adapt open blueprints into locally relevant solutions. Further, the technical achievement of the software upgrade developed by the workshop participants (see earlier) shows how open hardware approaches can simultaneously address technological needs while building local innovation capacity, a combination that proprietary "black-box" hardware prevents by restricting deep technical understanding and modifications. 

Our OpenDosimeter workshop thus demonstrates how combining open hardware with local stakeholder incentives creates a framework for developing adapted technical solutions while empowering next-generation innovators. The potential of this framework is perhaps best captured by one engineering student reflecting in the exit survey (Fig.~\ref{fig:4}c): \textit{"I have realized that people actually need these tools. And it’s crazy because I can make them."}  

\section*{Discussion}

OpenDosimeter represents a significant advancement in cost-effective personal dosimetry, suitable for distributed manufacturing and local calibration with direct access to dose exposure readings. Our results demonstrate that this open-source device offers performance comparable to commercial dosimeters that are >10$\times$ more expensive, in operational conditions relevant for occupational exposure in radiology departments. Most importantly, the open design allows maximum reproducibility on a global scale.

In particular, we show that by using $^{241}$Am for spectral calibration together with our transparent dose calculation approach, we can achieve surprisingly accurate values of the effective dose ($\pm$25\% within its dynamic range, cf. Fig.~\ref{fig:3}c). Furthermore, its real-time readout and data logging meet the standards set by state-of-the-art commercial active dosimeters at a component cost of <\$100. Despite these strengths, there are several areas where improvements are underway in upcoming upgrades of the OpenDosimeter:

\textbf{Power consumption:} The current version operates for up to 20 hours on a full charge, drawing <70 mA @ 3.7V with a 1200 mAh battery. While this is sufficient for 2--3 work days, there is significant room for improvement. By optimizing the choice of electronic components on the OpenDosimeter board, it should be easily possible to extend the battery life by at least >2$\times$.

\textbf{Integration:} To create an even more integrated and compact design, future versions will incorporate the microcontroller chip directly onto the OpenDosimeter board, eliminating the need for soldering a separate Raspberry Pi Pico. Furthermore, the recently released Pico 2 chips (e.g., RP2350) will improve overall performance at a negligible cost increase.

\textbf{Dose range:} The current reliable dose rate range, up to $\sim$1 mSv/h for energies up to 120 keV, can be sufficient for occupational X-ray exposure in most radiology departments. Some literature reports that typical scatter doses in radiography suites are in the range of 1--2 µSv/h \cite{onu2024scatter} (background levels: 0.1--0.5 µSv/h), detectable with the OpenDosimeter yet too low for most commercial equivalents (e.g., RaySafe i3 with its detection limit of $\sim$20 µSv/h from our studies). However, for applications where much higher instantaneous dose rates are expected, such as in interventional radiology, the current performance of OpenDosimeter is insufficient. During interventional procedures with fluoroscopy and radiographic acquisition, instantaneous exposure to the operators can reach 5 mSv/h and 50 mSv/h, respectively \cite{vano2023occupational}. To address this limitation, we are exploring two separate avenues for \textit{hardware} and \textit{software} modifications. 
    
On the \textit{hardware} side, we can improve the dose rate range by using a smaller area LYSO crystal to detect fewer counts at the same dose rates. One can also experiment with thinner LYSO crystals (e.g., 0.5 mm instead of 1 mm) or filtration (e.g., with Al or Cu), and smaller area SiPMs (e.g., 3$\times$3 mm or 1$\times$1 mm) to further reduce the overall count rate which will extend the linear range of dose rate measurements. Another benefit of this approach is that it will further reduce the cost. While this approach may reduce sensitivity at low dose rates, our current OpenDosimeter design already measures a count rate of $\sim$2000 cps at a relatively low dose rate of 10 µSv/h. This high count rate at low dose rates suggests that we have sufficient margin to map the dynamic range of the output count rate to a linear dose rate region with a higher maximum value before saturation is reached. With this approach, a 50$\times$ improvement of the maximum dose rate (to 50 mSv/h) should be possible. 
    
On the \textit{software} side, a promising approach is to continuously integrate the signal for a so-called "energy-integrating" dose calculation, running parallel to the current photon counting algorithm. This software-based solution aims to enable simultaneous probing of medium to high dose rates (1-50 mSv/h) without hardware changes. Our preliminary tests indicate that the amplified SiPM signal (cf. Fig.~\ref{fig:1}d) saturates beyond an instantaneous dose rate of 50 mSv/h, suggesting that this is a viable path forward.

Through our pilot workshop (Fig.~\ref{fig:4}) we demonstrate that open hardware can serve as an effective catalyst for sustainable local innovation when aligned with local stakeholder interests. By lowering barriers to entry through openly shared designs and documentation, we believe that projects like OpenDosimeter enable next-generation innovators to develop deep domain-specific technical competency. We propose that this approach to capacity building, combining technical education using open hardware with stakeholder incentives, could serve as a model for similar initiatives in other domains where accessible technology solutions are needed to address critical community needs.

To summarize, OpenDosimeter is to our knowledge the first open hardware dosimeter capable of accurate, real-time calculations of effective dose. Its innovative calibration procedure, utilizing $^{241}$Am found in ionization smoke detectors, enables local calibration without reliance on specialized facilities. The open design fosters reproducibility and local ownership, thereby supporting global capacity building in radiation safety. As a living project, future software and hardware iterations will continually enhance functionality. Moreover, the open-source licensing (GPLv3) encourages derivative works, which we hope will facilitate local commercialization to address the global demand for personal dosimeters.

\section*{Materials and Methods}
\vspace{-2mm}
\subsection*{Design, documentation, and reproducibility}
\vspace{-2mm}
The \href{https://github.com/OpenDosimeter/OpenDosimeter}{OpenDosimeter GitHub repository} contains comprehensive open-source documentation, including detailed instructions for both hardware assembly and software implementation. Additionally, video instructions referenced throughout the documentation can be found on the \href{https://www.youtube.com/@opendosimeter}{OpenDosimeter YouTube channel}. The data in Fig.~\ref{fig:1}c for the SiPM photon detection efficiency (PDE) is from the component datasheet (MICROFC-60035, Onsemi), while the LYSO emission profile is from the Luxium Solutions "LYSO Scintillation Material" datasheet. The data in Fig.~\ref{fig:1}d was acquired by probing the analog signals of an assembled OpenDosimeter (cf. "Debug pins" in Fig.~\ref{fig:1}a) using a 4-channel oscilloscope (DSOX3024G, Keysight).

\subsection*{Calibration and dose calculation}
\vspace{-2mm}
To extract samples of $^{241}$Am, we disassembled ionization smoke detectors (cf. Fig.~\ref{fig:2}a) from various brands (Model 5304, Universal Security Instruments; Model 9120BFF, First Alert; Model i9040, Kidde). While the activity seemed to vary slightly between models, they all performed well for our spectral calibration procedure. The data on the $^{241}$Am $\gamma$-emission in Fig.~\ref{fig:1}b (inset, green) is from the database accompanying our CdTe spectrometer (X-123CdTe, Amptek). The data in Fig.~\ref{fig:2}c is from the Luxium Solutions "LYSO Scintillation Material" datasheet. The data in Fig.~\ref{fig:2}d is from the ICRP 116 report (Table A.1, AP-direction) \cite{ICRP116}.

\subsection*{Characterization and performance demonstration}
\vspace{-2mm}
For our reference dose rate measurements, we used the RaySafe X2 system with the survey sensor, which is commonly used for reliable scatter dose survey measurements in radiology departments. X-ray experiments were conducted with a clinical X-ray tube operating with varying voltage (90 or 120 kV), current (20--200 mA, 10 steps) and Pb-filtration (2--8 mm). This resulted in average energies between 50-80 keV (reported by the RaySafe X2 system) depending on tube voltage and Pb-filtration thickness, incident on the devices (OpenDosimeter/RaySafe i3/RaySafe X2) placed 1 meter directly in front of the tube. By varying the Pb-filtration/voltage and sweeping over the tube current settings, we were able to gather data points over a range of reference dose rates from approximately 10 µSv/h to 5 mSv/h (Fig.~\ref{fig:3}b,c).

\subsection*{Pilot workshop for capacity-building} 
\vspace{-2mm}
We conducted the pilot workshop (Fig.~\ref{fig:4}) over seven days at Strathmore University, Nairobi, Kenya between the dates of Aug 28 to Sept 6, 2024. We selected twenty undergraduate engineering students (50/50 female/male ratio) initially, with twelve completing the full workshop (the remaining dropped out throughout the workshop period due to courses and other obligations). The workshop was held physically at the Makerspace Research Lab at the School of Computing and Engineering Sciences, which had the equipment needed to build the OpenDosimeter hardware (soldering stations, oscilloscopes, and other basic electronic tools). The student participants performed software programming of the devices using their own laptops. 

On the stakeholder side, we engaged with a director at the Kenya Nuclear Regulatory Authority (regulatory policy) who visited the makerspace during the workshop. The workshop curriculum also included visits to the Mama Lucy Kibaki Hospital (clinical radiology) as well as the Secondary Standards Dosimetry Laboratory at the Kenya Bureau of Standards (metrology).
\vspace{4mm}
\begin{acknowledgments}
The OpenDosimeter project is a derivative of the Open Gamma Detector project \cite{OpenGammaDetector}. We thank Shirin Pourashraf at Stanford University for assisting with SiPM configurations. We also thank the students who participated in the workshop at Strathmore University for their feedback on the OpenDosimeter design and its reproducibility. We acknowledge funding support from the Knut and Alice Wallenberg Foundation, and the King Center for Global Development at Stanford University.

\end{acknowledgments}

\section*{Author contributions}
\vspace{-2mm}
N.G. and K.S. conceived the project. N.G., A.K., A.S.W., and K.S. designed the device and its functionality. G.A. and E.A. gave input on functionality and user design. M.R. and K.S. developed the hardware and software. N.R.B., J.W., and K.S. performed the characterization. P.K. and K.S. organized the workshop at Strathmore University. N.R.B., P.K., and K.S. ran the workshop. K.S. analyzed the data. J.L. and K.S. wrote the open-source documentation. K.S. led the project and wrote the manuscript with input from all authors.

\bibliography{main}

%apsrev4-2.bst 2019-01-14 (MD) hand-edited version of apsrev4-1.bst
%Control: key (0)
%Control: author (8) initials jnrlst
%Control: editor formatted (1) identically to author
%Control: production of article title (0) allowed
%Control: page (0) single
%Control: year (1) truncated
%Control: production of eprint (0) enabled
\providecommand{\noopsort}[1]{}\providecommand{\singleletter}[1]{#1}%
\begin{thebibliography}{12}%
\makeatletter
\providecommand \@ifxundefined [1]{%
 \@ifx{#1\undefined}
}%
\providecommand \@ifnum [1]{%
 \ifnum #1\expandafter \@firstoftwo
 \else \expandafter \@secondoftwo
 \fi
}%
\providecommand \@ifx [1]{%
 \ifx #1\expandafter \@firstoftwo
 \else \expandafter \@secondoftwo
 \fi
}%
\providecommand \natexlab [1]{#1}%
\providecommand \enquote  [1]{``#1''}%
\providecommand \bibnamefont  [1]{#1}%
\providecommand \bibfnamefont [1]{#1}%
\providecommand \citenamefont [1]{#1}%
\providecommand \href@noop [0]{\@secondoftwo}%
\providecommand \href [0]{\begingroup \@sanitize@url \@href}%
\providecommand \@href[1]{\@@startlink{#1}\@@href}%
\providecommand \@@href[1]{\endgroup#1\@@endlink}%
\providecommand \@sanitize@url [0]{\catcode `\\12\catcode `\$12\catcode
  `\&12\catcode `\#12\catcode `\^12\catcode `\_12\catcode `\%12\relax}%
\providecommand \@@startlink[1]{}%
\providecommand \@@endlink[0]{}%
\providecommand \url  [0]{\begingroup\@sanitize@url \@url }%
\providecommand \@url [1]{\endgroup\@href {#1}{\urlprefix }}%
\providecommand \urlprefix  [0]{URL }%
\providecommand \Eprint [0]{\href }%
\providecommand \doibase [0]{https://doi.org/}%
\providecommand \selectlanguage [0]{\@gobble}%
\providecommand \bibinfo  [0]{\@secondoftwo}%
\providecommand \bibfield  [0]{\@secondoftwo}%
\providecommand \translation [1]{[#1]}%
\providecommand \BibitemOpen [0]{}%
\providecommand \bibitemStop [0]{}%
\providecommand \bibitemNoStop [0]{.\EOS\space}%
\providecommand \EOS [0]{\spacefactor3000\relax}%
\providecommand \BibitemShut  [1]{\csname bibitem#1\endcsname}%
\let\auto@bib@innerbib\@empty
%</preamble>
\bibitem [{\citenamefont {{International Atomic Energy
  Agency}}(2018)}]{IAEA2018}%
  \BibitemOpen
  \bibfield  {author} {\bibinfo {author} {\bibnamefont {{International Atomic
  Energy Agency}}},\ }\href@noop {} {\emph {\bibinfo {title}
  {\href{https://www.iaea.org/publications/11113/occupational-radiation-protection}{Occupational
  Radiation Protection}}}},\ \bibinfo {type} {Safety Standards Series}\
  \bibinfo {number} {GSG-7}\ (\bibinfo  {institution} {IAEA},\ \bibinfo
  {address} {Vienna},\ \bibinfo {year} {2018})\BibitemShut {NoStop}%
\bibitem [{\citenamefont {CAJOE}(2023)}]{CAJOE2023}%
  \BibitemOpen
  \bibfield  {author} {\bibinfo {author} {\bibnamefont {CAJOE}},\ }\href@noop
  {} {\bibinfo {title} {{RadiationD-v1.1}}},\ \bibinfo {howpublished}
  {\url{https://github.com/SensorsIot/Geiger-Counter-RadiationD-v1.1-CAJOE-}}
  (\bibinfo {year} {2023}),\ \bibinfo {note} {accessed: 2024-09-04}\BibitemShut
  {NoStop}%
\bibitem [{uRA(2017)}]{uRADMonitor}%
  \BibitemOpen
  \href@noop {} {\bibinfo {title} {{uRADMonitor}}},\ \bibinfo {howpublished}
  {\url{https://www.uradmonitor.com/tag/open-source/}} (\bibinfo {year}
  {2017}),\ \bibinfo {note} {accessed: 2024-09-04}\BibitemShut {NoStop}%
\bibitem [{\citenamefont {OpenGeiger}(2023)}]{OpenGeiger2023}%
  \BibitemOpen
  \bibfield  {author} {\bibinfo {author} {\bibnamefont {OpenGeiger}},\
  }\href@noop {} {\bibinfo {title} {{OpenGeiger}}},\ \bibinfo {howpublished}
  {\url{http://opengeiger.de/index_en.html}} (\bibinfo {year} {2023}),\
  \bibinfo {note} {accessed: 2024-09-04}\BibitemShut {NoStop}%
\bibitem [{\citenamefont {NuclearPhoenix}(2024)}]{OpenGammaDetector}%
  \BibitemOpen
  \bibfield  {author} {\bibinfo {author} {\bibnamefont {NuclearPhoenix}},\
  }\href@noop {} {\bibinfo {title} {{Open Gamma Detector}}},\ \bibinfo
  {howpublished}
  {\url{https://github.com/OpenGammaProject/Open-Gamma-Detector}} (\bibinfo
  {year} {2024}),\ \bibinfo {note} {accessed: 2024-09-04}\BibitemShut {NoStop}%
\bibitem [{\citenamefont {UniversalScientificTechnologies}(2024)}]{LABDOS01}%
  \BibitemOpen
  \bibfield  {author} {\bibinfo {author} {\bibnamefont
  {UniversalScientificTechnologies}},\ }\href@noop {} {\bibinfo {title}
  {{LABDOS01}}},\ \bibinfo {howpublished}
  {\url{https://github.com/UniversalScientificTechnologies/LABDOS01}} (\bibinfo
  {year} {2024}),\ \bibinfo {note} {accessed: 2024-09-04}\BibitemShut {NoStop}%
\bibitem [{\citenamefont {Gonz{\'a}lez}\ \emph {et~al.}(2016)\citenamefont
  {Gonz{\'a}lez}, \citenamefont {Aguilar}, \citenamefont {Conde}, \citenamefont
  {Hern{\'a}ndez}, \citenamefont {Moliner}, \citenamefont {Vidal},
  \citenamefont {S{\'a}nchez}, \citenamefont {S{\'a}nchez}, \citenamefont
  {Correcher}, \citenamefont {Molinos} \emph {et~al.}}]{gonzalez2016pet}%
  \BibitemOpen
  \bibfield  {author} {\bibinfo {author} {\bibfnamefont {A.~J.}\ \bibnamefont
  {Gonz{\'a}lez}}, \bibinfo {author} {\bibfnamefont {A.}~\bibnamefont
  {Aguilar}}, \bibinfo {author} {\bibfnamefont {P.}~\bibnamefont {Conde}},
  \bibinfo {author} {\bibfnamefont {L.}~\bibnamefont {Hern{\'a}ndez}}, \bibinfo
  {author} {\bibfnamefont {L.}~\bibnamefont {Moliner}}, \bibinfo {author}
  {\bibfnamefont {L.~F.}\ \bibnamefont {Vidal}}, \bibinfo {author}
  {\bibfnamefont {F.}~\bibnamefont {S{\'a}nchez}}, \bibinfo {author}
  {\bibfnamefont {S.}~\bibnamefont {S{\'a}nchez}}, \bibinfo {author}
  {\bibfnamefont {C.}~\bibnamefont {Correcher}}, \bibinfo {author}
  {\bibfnamefont {C.}~\bibnamefont {Molinos}}, \emph {et~al.},\ }\bibfield
  {title} {\bibinfo {title} {\href{https://doi.org/10.1109/TNS.2016.2522179}{A
  PET design based on SiPM and monolithic LYSO crystals: performance
  evaluation}},\ }\href@noop {} {\bibfield  {journal} {\bibinfo  {journal}
  {IEEE Transactions on Nuclear Science}\ }\textbf {\bibinfo {volume} {63}},\
  \bibinfo {pages} {2471} (\bibinfo {year} {2016})}\BibitemShut {NoStop}%
\bibitem [{\citenamefont {Melcher}(2000)}]{melcher2000scintillation}%
  \BibitemOpen
  \bibfield  {author} {\bibinfo {author} {\bibfnamefont {C.~L.}\ \bibnamefont
  {Melcher}},\ }\bibfield  {title} {\bibinfo {title}
  {\href{https://jnm.snmjournals.org/content/41/6/1051}{Scintillation crystals
  for PET}},\ }\href@noop {} {\bibfield  {journal} {\bibinfo  {journal}
  {Journal of Nuclear Medicine}\ }\textbf {\bibinfo {volume} {41}},\ \bibinfo
  {pages} {1051} (\bibinfo {year} {2000})}\BibitemShut {NoStop}%
\bibitem [{\citenamefont {Alva-S{\'a}nchez}\ \emph {et~al.}(2018)\citenamefont
  {Alva-S{\'a}nchez}, \citenamefont {Zepeda-Barrios}, \citenamefont
  {D{\'\i}az-Mart{\'\i}nez}, \citenamefont {Murrieta-Rodr{\'\i}guez},
  \citenamefont {Mart{\'\i}nez-D{\'a}valos},\ and\ \citenamefont
  {Rodr{\'\i}guez-Villafuerte}}]{alva2018understanding}%
  \BibitemOpen
  \bibfield  {author} {\bibinfo {author} {\bibfnamefont {H.}~\bibnamefont
  {Alva-S{\'a}nchez}}, \bibinfo {author} {\bibfnamefont {A.}~\bibnamefont
  {Zepeda-Barrios}}, \bibinfo {author} {\bibfnamefont {V.}~\bibnamefont
  {D{\'\i}az-Mart{\'\i}nez}}, \bibinfo {author} {\bibfnamefont
  {T.}~\bibnamefont {Murrieta-Rodr{\'\i}guez}}, \bibinfo {author}
  {\bibfnamefont {A.}~\bibnamefont {Mart{\'\i}nez-D{\'a}valos}},\ and\ \bibinfo
  {author} {\bibfnamefont {M.}~\bibnamefont {Rodr{\'\i}guez-Villafuerte}},\
  }\bibfield  {title} {\bibinfo {title}
  {\href{https://doi.org/10.1038/s41598-018-35684-x}{Understanding the
  intrinsic radioactivity energy spectrum from $^{176}$Lu in LYSO/LSO
  scintillation crystals}},\ }\href@noop {} {\bibfield  {journal} {\bibinfo
  {journal} {Scientific reports}\ }\textbf {\bibinfo {volume} {8}},\ \bibinfo
  {pages} {17310} (\bibinfo {year} {2018})}\BibitemShut {NoStop}%
\bibitem [{\citenamefont {{International Commission on Radiological
  Protection}}(2010)}]{ICRP116}%
  \BibitemOpen
  \bibfield  {author} {\bibinfo {author} {\bibnamefont {{International
  Commission on Radiological Protection}}},\ }\bibfield  {title} {\bibinfo
  {title} {\href{https://www.icrp.org/publication.asp?id=ICRP Publication
  116}{Conversion Coefficients for Radiological Protection Quantities for
  External Radiation Exposures}},\ }\href@noop {} {\bibfield  {journal}
  {\bibinfo  {journal} {Annals of the ICRP}\ }\textbf {\bibinfo {volume} {40}}
  (\bibinfo {year} {2010})},\ \bibinfo {note} {{ICRP Publication
  116}}\BibitemShut {NoStop}%
\bibitem [{\citenamefont {Onu}\ and\ \citenamefont
  {Nzotta}(2024)}]{onu2024scatter}%
  \BibitemOpen
  \bibfield  {author} {\bibinfo {author} {\bibfnamefont {C.~P.}\ \bibnamefont
  {Onu}}\ and\ \bibinfo {author} {\bibfnamefont {C.~C.}\ \bibnamefont
  {Nzotta}},\ }\bibfield  {title} {\bibinfo {title}
  {\href{https://doi.org/10.1016/j.apradiso.2024.111472}{Scatter radiation
  levels in X-ray rooms during chest radiography}},\ }\href@noop {} {\bibfield
  {journal} {\bibinfo  {journal} {Applied Radiation and Isotopes}\ ,\ \bibinfo
  {pages} {111472}} (\bibinfo {year} {2024})}\BibitemShut {NoStop}%
\bibitem [{\citenamefont {Vano}\ \emph {et~al.}(2023)\citenamefont {Vano},
  \citenamefont {Fernandez-Soto}, \citenamefont {Ten},\ and\ \citenamefont
  {Sanchez~Casanueva}}]{vano2023occupational}%
  \BibitemOpen
  \bibfield  {author} {\bibinfo {author} {\bibfnamefont {E.}~\bibnamefont
  {Vano}}, \bibinfo {author} {\bibfnamefont {J.~M.}\ \bibnamefont
  {Fernandez-Soto}}, \bibinfo {author} {\bibfnamefont {J.~I.}\ \bibnamefont
  {Ten}},\ and\ \bibinfo {author} {\bibfnamefont {R.~M.}\ \bibnamefont
  {Sanchez~Casanueva}},\ }\bibfield  {title} {\bibinfo {title}
  {\href{https://doi.org/10.1259/bjr.20220607}{Occupational and patient doses
  for interventional radiology integrated into a dose management system}},\
  }\href@noop {} {\bibfield  {journal} {\bibinfo  {journal} {The British
  Journal of Radiology}\ }\textbf {\bibinfo {volume} {96}},\ \bibinfo {pages}
  {20220607} (\bibinfo {year} {2023})}\BibitemShut {NoStop}%
\end{thebibliography}%

\end{document}